\documentclass[preprint,12pt]{elsarticle}

\usepackage{amssymb}

\journal{Nuclear Physics A}

\begin{document}

\begin{frontmatter}



\title{Shear oscillations in the hadron-quark mixed phase}


\author{Hajime Sotani$^1$} 
\author{Toshiki Maruyama$^2$}
\author{Toshitaka Tatsumi$^3$}

\address{
$^1$Yukawa Institute for Theoretical Physics, Kyoto University, Kyoto 606-8502, Japan\\
$^2$Advanced Science Research Center, Japan Atomic Energy Agency, Tokai, Ibaraki 319-1195, Japan\\
$^3$Department of Physics, Kyoto University, Kyoto 606-8502, Japan
}

\begin{abstract}
Inside neutron stars, the hadron-quark mixed phase is expected during
the first order phase transition from the hadron phase to the quark phase. 
The geometrical structure of the mixed phase strongly depends on the surface tension at the hadron-quark interface. 
We evaluate the shear modulus which is one of the specific properties of the hadron-quark mixed phase. 
As an application, we study shear oscillations due to the hadron-quark mixed phase in neutron stars. 
We find that the frequencies of shear oscillations depend strongly on the surface tension;
with a fixed stellar mass, the fundamental frequencies are almost proportional to the surface tension. 
Thus, one can estimate the value of surface tension via the observation of stellar oscillations 
with the help of the information on the stellar mass.
\end{abstract}


\end{frontmatter}


\section{Introduction}
\label{sec:I}

Neutron stars are formed in supernova explosions, which arise at the final stage of the evolution of massive stars. The density can exceed the standard nuclear density of $\rho_0\approx 0.17$ fm$^{-3}$ inside neutron stars. Since such a high density is almost impossible to be realized on the Earth, neutron stars may be a unique ``laboratory" to investigate the properties of matter around and beyond the nuclear density. One of the possible ways to see the properties of neutron star matter could be the observations of gravitational waves emitted from neutron stars. Since the gravitational waves with high permeability will bring us raw information on the wave sources, one can directly see the properties of neutron star matter. 
Thus observations of gravitational waves will provide us with the astrophysical data to reveal basic properties of dense matter (e.g., \cite{AK1996,AK1998,KS1999,AC2001,Sotani2001,Sotani2003,BFG2005,Erich2011}), and to examine the theory of strong gravity (e.g., \cite{Sotani2004,Sotani2009a,Sotani2010,BPPL2012}). 
Now, the worldwide projects are going on to detect gravitational waves  associated with the astrophysical phenomena involving compact objects  \cite{Barish2005,LIGO1,LIGO2}. Another way to see the properties of neutron-star matter could be the direct observations of global oscillations of neutron stars. One would know then the stellar mass, radius, and equation of state (EOS) from such observations. This method is often referred to asteroseismology, which is quite similar to the helioseismology. 

Unlike gravitational waves, fortunately the observational evidences of neutron-star oscillations have been detected, i.e., the quasi-periodic oscillations (QPOs) in the X-ray afterglow of the giant flares in soft gamma repeaters (SGRs). 
Up to now, at least three giant flares have been detected in SGR 0526-66, SGR 1900+14, and SGR 1806-20. 
Furthermore, through the timing analysis of the X-ray afterglow in those giant flares, the specific QPO frequencies have been also extracted in the range from tens Hz up to a few kHz \cite{WS2006}. 
Since the central objects in SGRs are considered to be magnetars which are strongly magnetized neutron stars, the discovered QPOs in giant flares could be due to neutron star oscillations. In order to explain these QPO frequencies theoretically, a lot of numerical attempts have been done not only by the shear oscillations in the crust of neutron stars but also by the magnetic oscillations \cite{Levin2006,Lee2007,SA2007,Sotani2007,Sotani2008a,Sotani2008b,Sotani2009,CBK2009,CSF2009,GCFMS2011,CK2011}. In addition, ascribing the observed QPOs to shear oscillations in the crust of neutron stars, the possibilities are also pointed out to reveal the properties of inhomogeneous nuclear matter in the crust \cite{SW2009,Sotani2011,GNHL2011,Sotani2012,Sotani2012b}.

The structure of a neutron star is considered as follows: ocean of liquid iron exists in the vicinity of stellar surface up to the density $\sim 10^6-10^8$ g cm$^{-3}$, subsequently the crust region exists from the bottom of iron ocean up to the density of the order of $\rho_0$. Then a fluid core exists in higher density region. 
In most part of the crust, nuclei form a bcc lattice due to the Coulomb interaction, while the existence of exotic nuclear shapes at the bottom of crust is also suggested in the recent studies \cite{Ravenhall83,Hashimoto84,Lorenz1993,Oyamatsu1993,SOT1995, maru2005,new,oka}. According to such studies, with increasing density, the shape of nuclear matter changes from sphere (bcc lattice\footnote{
Very recently, we and our collaborators have found that fcc lattice of spherical nuclei can be the ground state at some densities,
choosing the optimum sizes of the cell and nuclei as well as the inhomogeneous electron distribution \cite{oka}.
}), to cylinder, slab, cylindrical hole, spherical bubble, and uniform matter (inner fluid core), which is collectively called ``nuclear pasta." 

On the other hand, there are still many uncertainties in the core region. 
For example, hyperons  appear in beta-equilibrium  nuclear matter when the density becomes higher than $\sim 2-3\rho_0$, or non-hadronic quark matter might exist in the innermost stellar core \cite{rev,rev1}. 
Depending on the presence of these exotic components, structure of neutron stars dramatically changes \cite{BBSS2002,Maruyama2007}. 
As for the hadron-quark (HQ) phase transition, there are further uncertainties such as EOS of quark matter and/or the deconfinement mechanism. Anyhow the HQ mixed phase may emerge as a consequence of the Gibbs conditions, supposing that the HQ phase transition is of the first order \cite{gle}.

The HQ mixed phase looks like the nuclear pasta and strongly depends on the Coulomb interaction and the surface tension at the hadron-quark interface, which are called ``fine-size effects" \cite{hei,vos}. Namely, whether non-uniform (pasta) structures appear in the HQ mixed phase is subject to a balance between the surface tension and the Coulomb repulsion \cite{Maruyama2007,Yasutake2009}, which is similar to the situation in the crust \cite{maru2005} and kaon condensed matter \cite{Maruyama2006}. However, to determine the surface tension with the experiments on the Earth is quite difficult because temperature becomes too high for the HQ mixed phase to appear during the relativistic heavy-ion collisions. One of the possibilities to distinguish the finite-size effects on the HQ mixed phase in neutron stars might be the observations of astronomical phenomena. Actually, we suggested such a possibility via direct observations of gravitational waves emitted from neutron stars with the HQ mixed phase \cite{SYMT2011}. Such attempts are very challenging, but there are very few literatures.

On the other hand, shear modulus is one of the specific properties of the HQ mixed phase due to the existence of non-uniform structures. 
In fact,  the shear modulus becomes zero without non-uniform structures. 
As mentioned before, however, the properties of such a phase are quite uncertain since the deconfinement mechanism is not clarified. Hence, this article aims mainly at giving the shear modulus and the shear speed in the HQ mixed phase in order to get an insight of the HQ mixed phase. 
Additionally, as an application, we also explore shear oscillations to see the dependence on the surface tension. 
We examine the frequencies of shear oscillations within the relativistic Cowling approximation, modeling neutron stars with the HQ mixed phase.

This article is organized as follows: In Sec.\ \ref{sec:II}, we describe the equilibrium of neutron stars and the adopted EOS, where we will discuss the shear modulus in the HQ mixed phase. In Sec.\ \ref{sec:III},  we show equations governing shear oscillations and the boundary conditions to determine the eigenfrequencies. Additionally, the obtained spectra of such oscillations will be shown. At the end, we make a conclusion in Sec.\ \ref{sec:IV}. We adopt the unit of $c=G=1$ in this article, where $c$ and $G$ denote the speed of light and the gravitational constant, respectively, and the metric signature is $(-,+,+,+)$.

\section{Neutron Star Models}
\label{sec:II}

In this article, we focus on non-rotating neutron stars with the HQ mixed phase. The equilibrium configuration of such relativistic objects is given by spherically symmetric solutions of the Tolman-Oppenheimer-Volkoff (TOV) equations. In this situation, the metric can be expressed as
\begin{equation}
  ds^2 = -e^{2\Phi}dt^2 +  e^{2\Lambda}dr^2 + r^2 (d\theta^2 + \sin^2\theta d\phi^2),
\end{equation}
where $\Phi$ and $\Lambda$ are functions with respect to radial coordinate $r$. Mass function $m(r)$ is associated with the metric function $\Lambda$ as $m(r)=r(1- e^{-2\Lambda})/2$, which satisfies
\begin{equation}
  m(r)' = 4\pi r^2 \varepsilon(r),
\end{equation}
where the prime on variables denotes the partial derivative with respect to $r$, and $\varepsilon(r)$ is the energy density. The distributions of pressure $p(r)$ and metric function $\Phi(r)$ can be determined by solving the TOV equations;
\begin{eqnarray}
  p(r)' &=& -(\varepsilon + P)\Phi', \\
  \Phi(r)' &=& \frac{m+4\pi r^3 p}{r(r-2m)}.
\end{eqnarray}
In addition to these equations, one needs to prepare the EOS to close the coupled equations.

According to Ref. \cite{Maruyama2007}, we especially adopt the EOS including hyperons with the HQ mixed phase, which properly takes into account the finite-size effects. That is, the EOS for hadron phase is adopted the non-relativistic Brueckner-Hartree-Fock (BHF) EOS, while the EOS for quark matter is assumed a generalized phenomenological MIT bag model, where we assume that $u$ and $d$ quarks are massless while $s$ quarks have the mass of $m_{\rm s}=150$ MeV as in Refs. \cite{Maruyama2007,SYMT2011}. Depending on the surface tension at the HQ interface, the non-uniform structure could appear in the HQ mixed phase as a consequence of the Gibbs conditions, where the geometrical shapes of droplet, rod, slab, tube, and bubble are considered. The knowledge of the value of surface tension is very poor, but that value is theoretically estimated around $\sigma\approx 5-100$ MeV fm$^{-2}$ \cite{FJ1984,KKR1991,PF2010,PKR2012}. Since it is difficult to produce the HQ mixed phase with a larger value of $\sigma$, we especially adopt $\sigma=10$, 20, and 40 MeV fm$^{-2}$ in this article. In fact, the structure in the HQ mixed phase with $\sigma \approx 40-100$ MeV fm$^{-2}$ is almost the same as that with $\sigma =40$ MeV fm$^{-2}$ \cite{Maruyama2007}. Finally, for the lower density region, the above EOS should be connected to the hadronic EOS proposed by Negele and Vautherin \cite{EOS_NV}.

\begin{figure}[htbp]
\begin{center}
\includegraphics[scale=0.5]{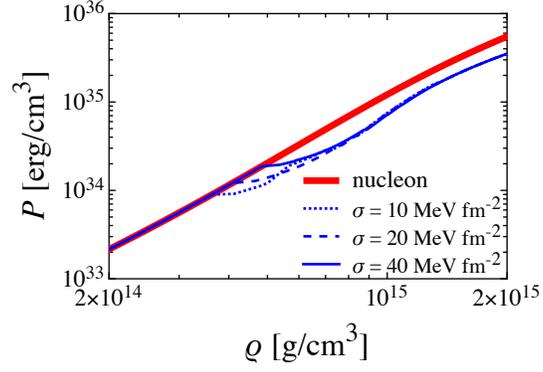} 
\end{center}
\caption{Relationship between the total energy density and the pressure.
}
\label{fig:eos}
\end{figure}

\begin{figure}[htbp]
\begin{center}
\includegraphics[scale=0.5]{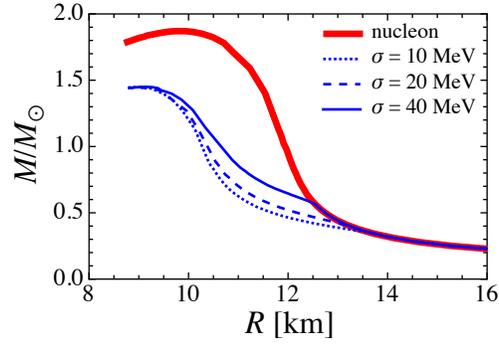} 
\end{center}
\caption{Stellar mass as a function of radius.
}
\label{fig:MR}
\end{figure}

The EOS for higher density region adopted in this article can be shown in Fig. \ref{fig:eos}, where the EOS composed of only nucleons is also shown for comparison. Then, as shown in Fig. \ref{fig:MR}, the stellar properties are obtained by solving the TOV equations. We should remark that the EOS adopted in this article can not reach the observed maximum mass, which is $M_{\rm max}=1.97\pm0.04M_\odot$ \cite{2M}, because introduction of quark matter usually makes EOS soft. Although some EOSs with quark matter whose maximum mass becomes over $M_{\rm max}$ have been suggested recently (e.g., \cite{BS2012,MHT2012}), there is no EOS dealing with the pasta structure in the HQ mixed phase. However, as mentioned before, the existence of pasta structure is essential to consider the shear properties. Thus, we adopt the above EOS in this article in order to see the dependence of shear modulus on the surface tension. Even if one adopts other EOSs, we believe that our result could be robust as far as the HQ mixed phase appears in the similar density region.

%

\begin{figure}[htbp]
\begin{center}
\includegraphics[scale=0.5]{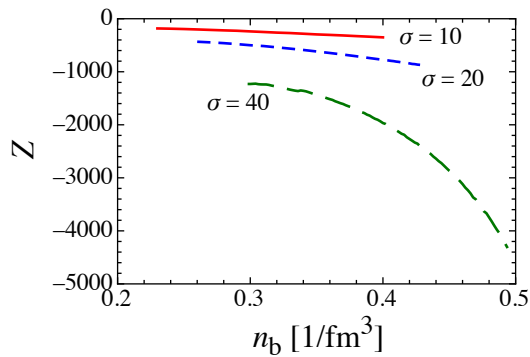} 
\end{center}
\caption{Total charge number $Z$ included in the quark spherical droplet.
}
\label{fig:ZZ}
\end{figure}

\begin{figure}[htbp]
\begin{center}
\includegraphics[scale=0.5]{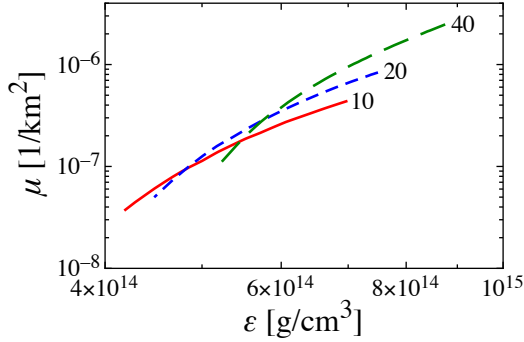} 
\end{center}
\caption{Shear modulus $\mu$ in the droplet region of the HQ mixed phase as a function of energy density, where the labels denote the corresponding values of $\sigma$ in the unit of MeV fm$^{-2}$.
}
\label{fig:shear}
\end{figure}

Although there are not so many literatures about the shear modulus in neutron stars, the formula about the shear modulus in neutron star crust has been presented in the zero temperature limit as
\begin{equation}
 \mu = 0.1194\times \frac{n_i (Ze)^2}{a}, \label{eq:shear}
\end{equation}
where $n_i$, $Ze$, and $a$ denote the ion number density, the ion charge, and the average ion spacing defined as $a^3=3/(4\pi n_i)$ \cite{OI1990,SHOII1991}. This formula is derived from the Monte Carlo calculations with the assumption that the shear modulus is averaged over all directions and ion forms a perfect bcc lattice. In order to apply the expression of the shear modulus (\ref{eq:shear}) for the droplet region in the HQ mixed phase, we consider that $n_i$ should be the number density of quark spherical droplets in the hadron sea, while $Ze$ should be the total charge included in the quark spherical droplet. In particular, we derive the total charge number $Z$ in the quark droplet by using the equation as
\begin{eqnarray}
    Z &=& \left(n_q - n_e\right)V_{\rm drop}, \\
    n_q &\equiv&  \frac{2n_u - n_d - n_s}{3},
\end{eqnarray}
where $n_u$, $n_d$, and $n_s$ are the number density of $u$, $d$, and $s$ quarks in the volume of the quark droplet $V_{\rm drop}$, $n_e$ is the electron number density in the Wigner-Seitz cell, and $n_q$ denotes the charge density of quarks inside the droplet in the HQ mixed phase. The calculated $Z$ is shown in Fig. \ref{fig:ZZ} as a function of the baryon number density $n_b$. We remark that the charge number in the quark droplet becomes negative, because the hadronic sea is mainly composed of proton and neutron. From Fig. \ref{fig:ZZ}, we can see that the charge number included in the quark droplet in the HQ mixed phase depends strongly on the surface tension $\sigma$ and the absolute value can become $10-100$ times larger than that in the crust nuclei, because the typical charge number in the crust nuclei is $\sim 40$ \cite{DH} (also see Appendix A for the simple estimation). Then, by using Eq. (\ref{eq:shear}), we calculate the shear modulus in the quark droplet region of the HQ mixed phase. We show it in Fig. \ref{fig:shear} as a function of energy density $\varepsilon$, where the labels in the figure denote the corresponding values of $\sigma$ in the unit of MeV fm$^{-2}$. As a result, we find that the shear modulus in the HQ mixed phase becomes about $10^3$ times larger than that in the crust region, considering that the typical value of $\mu$ in the crust region is around $\mu\simeq 10^{-10}-10^{-9}$ km$^{-2}$ \cite{Sotani2011}
\footnote{
In this paper, we examine the torsional oscillations in the HQ mixed phase. However, it would be also interesting to consider the shear modulus used in modeling the crystalline quark matter in the color superconducting phase \cite{PP1998,JO2012,Alford2008}.
}. We remark that the reason why the shear modulus in the HQ mixed phase is so much larger than that in the crust region is due to the difference of charge number in each phase as mentioned in Appendix A, in addition to the increase of baryon number density. Meanwhile, another specific property due to the HQ mixed phase is the shear speed defined as $v_s^2=\mu/(\rho+p)$. Roughly speaking, with $\sim10^3$ times larger shear modulus and with $\sim 10$ times larger baryon number density than those in the crust region, one can expect that the shear speed in the HQ mixed phase can be $\sim 10$ times larger than that in the crust region. In other words,  the propagation time in the HQ mixed phase with the shear speed becomes $\sim 10$ times smaller than that in the crust region. These estimations are important to consider the shear oscillations as shown in the below.

Finally, because of little knowledge concerning the shear modulus in the other pasta structures except for the spherical droplet phase, as a first step, we assume that $\mu=0$ except for the droplet region in this article as in Refs. \cite{GNHL2011,Sotani2012}. In fact, the elasticity in such region is expected to be lower than that in the droplet region \cite{PP1998}\footnote{
Recently, Johnson-McDaniel and Owen suggested that, depending on the surface tension, the effective shear moduli for the non-droplet pasta structures can be larger than that for the droplet phase \cite{JO2011}}.

%

%

\section{Torsional Oscillations}
\label{sec:III}

The stellar oscillations in the spherically symmetric stars can be classified into two families with their parities. The oscillations with polar parity accompany the density variation and stellar deformation, while those with axial parity are incompressible motion. Thus, one can examine the axial oscillations with high accuracy even if one neglects the metric perturbations by setting $\delta g_{\mu\nu}=0$, which is known as the relativistic Cowling approximation. With this approximation, the restoring force of the axial oscillations is the shear stress characterized by shear modulus $\mu$. Such oscillations are referred as shear oscillations and can be described by a single perturbation variable ${\cal Y}$, which is corresponding to the angular displacement in the $\phi$ direction. The variable ${\cal Y}$ is associated with the $\phi$-component of the perturbation of fluid four-velocity as
\begin{equation}
  \delta u^\phi =  e^{-\Phi}\partial_t{\cal Y}(t,r) \frac{1}{\sin\theta}\partial_\theta P_\ell(\cos\theta),
\end{equation}
where $P_\ell$ denotes the $\ell$-th order Legendre polynomial. The perturbational equation governing the shear oscillations can be derived from the linearized equation of motion \cite{ST1983}. Assuming ${\cal Y}(t,r)= e^{ i\omega t}{\cal Y}(r)$, such a perturbational equation can be written as
\begin{equation}
  {\cal Y}'' + \left[\left(\frac{4}{r}+\Phi'-\Lambda'\right)+\frac{\mu'}{\mu}\right]{\cal Y}' 
    + \left[\frac{\varepsilon+p}{\mu}\omega^2 e^{-2\Phi}-\frac{(\ell+2)(\ell-1)}{r^2}\right] e^{2\Lambda}{\cal Y} = 0.
  \label{eq:perturbation}
\end{equation}
Then, the problem to solve is reduced to the eigenvalue problem, imposing the appropriate boundary conditions. As in Refs. \cite{Sotani2011,Sotani2012,ST1983}, we impose zero-traction conditions at the both boundaries of the HQ mixed phase.
In practice, the matter element outside the HQ mixed phase can not affect the motion of torsional oscillations, which is the same situation in the crust torsional oscillations at the boundary between the crust and core regions. In this paper, we examine the torsional oscillations with a linear analysis. This means that we can calculate the frequencies of oscillations, while we can not say anything about the amplitude of oscillations. Consequently, it is also impossible to mention about the detectability of emitted gravitational waves. However, in general, the energy of emitted gravitational waves with axial oscillations is quite small, because such oscillations do not involve the density variation as mentioned above. So, it might be difficult to probe the properties of the HQ mixed phase by way of the direct observation of gravitational waves due to the torsional oscillations.

\begin{figure}[htbp]
\begin{center}
\includegraphics[scale=0.5]{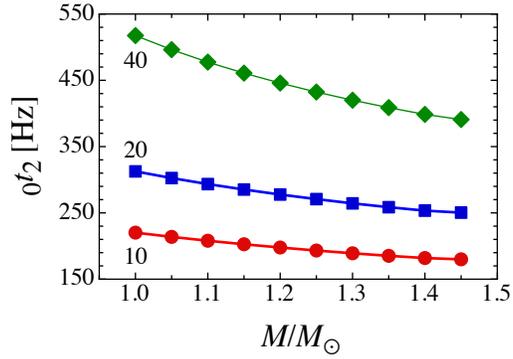} 
\end{center}
\caption{Frequencies of fundamental shear oscillations with $\ell=2$ as a function of the stellar mass, where the labels correspond to the values of $\sigma$ in the unit of MeV fm$^{-2}$.
}
\label{fig:0t2}
\end{figure}

\begin{figure}[htbp]
\begin{center}
\includegraphics[scale=0.5]{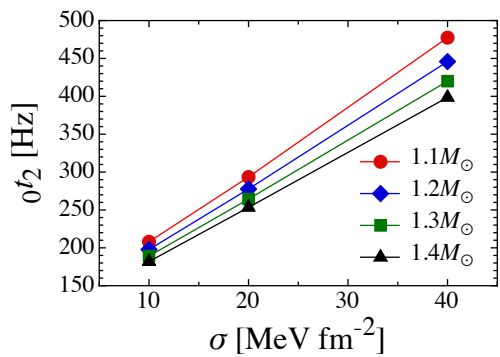} 
\end{center}
\caption{Frequencies of fundamental shear oscillations with $\ell=2$ as a function of surface tension for different values of stellar mass.
}
\label{fig:0t2-sigma}
\end{figure}

Before doing the numerical calculations, we can make a simple estimation about the frequencies of shear oscillations in the HQ mixed phase. Since the propagation time with shear speed in the HQ mixed phase becomes $\sim 10$ times smaller than that in the crust region, one can estimate that the frequencies of shear oscillations in the HQ mixed phase could become roughly $10$ times as large as those in the crust region. On the other hand, for $\sigma=10$, 20, and 40 MeV fm$^{-2}$, the calculated frequencies of fundamental shear oscillations with $\ell=2$ in the HQ mixed phase are shown in Fig. \ref{fig:0t2} as a function of the stellar mass. Considering that such frequencies in the crust region are around tens Hz, as in the above estimation, the frequencies of fundamental shear oscillations in the HQ mixed phase can become ten times larger than those in the crust region. Additionally, from Fig. \ref{fig:0t2}, one can observe that the frequencies of fundamental shear oscillations depend strongly on the value of $\sigma$. In practice, the frequencies for $\sigma=20$ and 40 MeV fm$^{-2}$ are $\sim$40\% and $\sim$120 \% larger than those for $\sigma=10$ MeV fm$^{-2}$. Thus, if one would identify the observed frequencies as the shear oscillations in the HQ mixed phase, one might be able to probe the properties of such exotic structure. Moreover, with fixed stellar mass, we plot the dependence of the frequencies of fundamental shear oscillations on the surface tension in Fig. \ref{fig:0t2-sigma}. From this figure, one can find the frequencies are almost proportional to $\sigma$. Thus, with the help of the other observation of stellar mass, one could put a constraint on the value of $\sigma$ by way of the observation of the frequency of fundamental shear oscillations.

%

\begin{figure}[htbp]
\begin{center}
\includegraphics[scale=0.5]{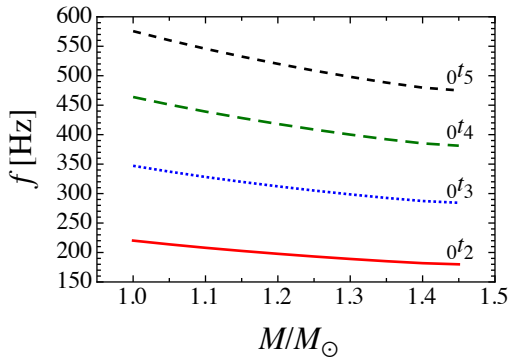} 
\end{center}
\caption{For $\sigma=10$ MeV fm$^{-2}$, frequencies of fundamental shear oscillations with $\ell=2$, 3, 4, and 5 are shown as a function of the stellar mass.
}
\label{fig:s10Mf}
\end{figure}

On the adopted stellar model, the shear oscillations with different values of $\ell$ can also exist as well as the $\ell=2$ oscillation. In order to see the behavior of such oscillations, for the stellar model with $\sigma=10$ MeV fm$^{-2}$, we plot the calculated fundamental frequencies with $\ell=2$, 3, 4, and 5 in Fig. \ref{fig:s10Mf} as a function of stellar mass, where the labels of ${}_0t_\ell$ are corresponding to the frequencies of $\ell$-th order oscillations. Note that the different combinations of $\ell$ and $\sigma$ may give the same frequency; for example, ${}_0t_2$ with large value of $\sigma$ might coincide with ${}_0t_3$ with small value of $\sigma$. As a result, the identification of the shear oscillation from one observation with the specific frequency might be difficult, even if one would know the mass of the source object. However, if one will simultaneously observe several oscillation frequencies from the object whose mass is known, one could be possible to put a constraint on $\sigma$ and to identify $\ell$ in such a way to explain the observed evidences all together. Probably, with the development of observation technology, such observations will become possible and we will see the properties of the HQ mixed phase by way of the observation of the stellar oscillations.

\begin{figure}[htbp]
\begin{center}
\includegraphics[scale=0.5]{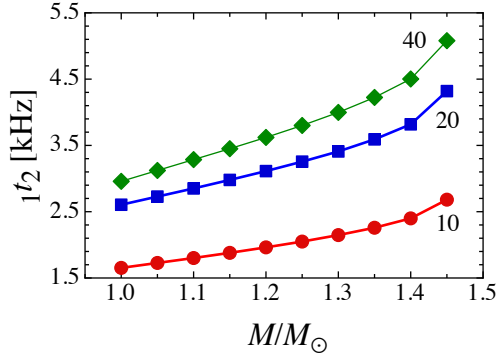} 
\end{center}
\caption{Frequencies of 1st overtones of shear oscillations with $\ell=2$ as a function of the stellar mass, where the labels correspond to the values of $\sigma$ in the unit of MeV fm$^{-2}$.
}
\label{fig:1t2}
\end{figure}

\begin{figure}[htbp]
\begin{center}
\includegraphics[scale=0.5]{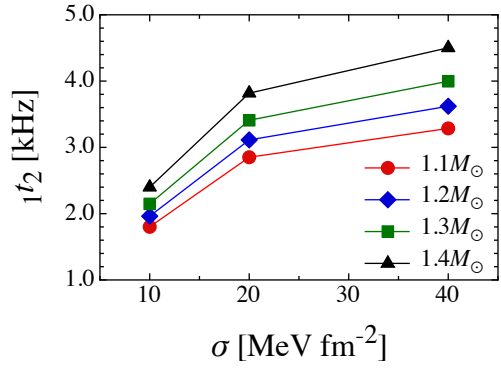} 
\end{center}
\caption{Frequencies of 1st overtones of shear oscillations with $\ell=2$ as a function of the surface tension with different values of stellar mass.
}
\label{fig:1t2-sigma}
\end{figure}

Finally, the frequencies of 1st overtone of shear oscillations are shown in Fig. \ref{fig:1t2}. One can see again that the frequencies in the HQ mixed phase becomes around ten times larger than those in the crust region due to the large shear speed in the HQ mixed phase. Additionally, one can see the strong dependence of frequencies on the surface tension, where the frequencies with $\sigma=20$ and 40 MeV fm$^{-2}$ become $\sim$58\% and $\sim$85\% larger than those with $\sigma=10$ MeV fm$^{-2}$. Compared with the fundamental oscillations, the dependence of frequencies of 1st overtone might not be so strong, but this difference could be still observable. It should be emphasized that the dependence of frequencies of 1st overtone on $\sigma$ is different from that of fundamental oscillations, i.e., the frequencies of 1st overtone with the fixed stellar mass are not proportional to the surface tension (see Fig. \ref{fig:1t2-sigma}). Thus, one might be able to put a severer constraint on the surface tension through both observations of frequencies of fundamental and overtone shear oscillations.

%

%

\section{Conclusion}
\label{sec:IV}

In this article, we have focused on the hadron-quark (HQ) mixed phase, which can appear inside neutron stars if hadron matter makes a phase transition into quark matter. Depending on the surface tension at the HQ interface, the non-uniform structures can appear in the HQ mixed phase, which produces the shear properties. With the EOS involving the non-uniform structure in the HQ mixed phase, we have estimated the shear modulus in this article. Then, we have found that the shear modulus depends strongly on the surface tension, which becomes $\sim 10^3$ times larger than that in the crust region. Probably, this is caused by the difference of the charge number including in the droplet.

Meanwhile, as an application, we have calculated the shear oscillations in the HQ mixed phase. 
As a result, we have found that the frequencies of shear oscillation in the HQ mixed phase could be around ten times larger than that in the crust region and those frequencies depend strongly on the value of the surface tension for the HQ interface. 
We also found that the frequencies of fundamental oscillations with the fixed stellar mass are almost proportional to the surface tension. Thus, with the help of the observation about the stellar mass, one might be able to determine the value of surface tension with using the observations of the frequencies of shear oscillations in the HQ mixed phase.

Finally, the resulting frequencies of fundamental oscillations in the HQ mixed phase are order of 100 Hz. This means that some of the QPO frequencies observed in giant flares, for example 150 Hz and even 626.5 Hz in SGR 1806-20 or 155 Hz in SGR 1900+14 \cite{WS2006}, might be associated with the shear oscillations in the HQ mixed phase. 
Although it has not been successful to get the collective view about the observed QPOs yet, the consideration of the shear oscillations in the HQ mixed phase may be able to solve the puzzle for the theoretical explanation of the QPO frequencies observed in giant flares.

H.S. is grateful to N.~Yasutake for his warm hospitality and fruitful discussions. 
This work was supported in part by Grants-in-Aid for Scientific Research on Innovative 
Areas through No.\ 23105711, No.\ 24105001, and No.\ 24105008 from MEXT, by Grant-in-Aid for Young Scientists (B) through No.\ 24740177 from JSPS, by the Yukawa International Program for Quark-hadron Sciences, and by the Grant-in-Aid for the global COE program ``The Next Generation of Physics, Spun from Universality and Emergence" from MEXT.


\appendix
\section{HQ pasta and the nuclear pasta}
\label{sec:app}

In this appendix A, we compare the properties associated with the shear modulus for the droplet phase in the nuclear pasta and the HQ pasta. In particular, in order to see the physical contributions clearly, we adopt the compressible liquid drop model in this appendix A \cite{rev1}. Now, we define the volume fraction, $w$, the electron fraction, $Y_e$, and the proton fraction, $Y_p$, as $w\equiv (R/R_{\rm W})^3$, $Y_e\equiv n_e/n_B$, and $Y_p\equiv n_p/n_B$, where $R$, $R_{\rm W}$, $n_e$, $n_p$, and $n_B$ are corresponding to the radius of droplet, the radius of Wigner-Seitz cell, the electron number density, the proton number density, and the average baryon number density. It is noted that one should consider $n_p$ for the nuclear pasta as the proton number density inside the droplet, while $n_p$ for the HQ pasta as that outside the droplet. Generally, the pasta structures are determined by the balance between the Coulomb and surface energies. Then, the Coulomb energy, $E_{\rm Coul}$, and the surface energy, $E_{\rm surf}$, per each volume of the Wigner-Seitz cell, $V$, in the nuclear pasta can be written as
\begin{eqnarray}
  \frac{E_{\rm Coul}^N}{V_N} &=& \frac{1}{w_N}\frac{4\pi}{5}\left(Y_e^N\right)^2\left(n_B^N\right)^2e^2R_N^2f_3(w_N),
     \label{eq:a1} \\
  \frac{E_{\rm surf}^N}{V_N} &=& \frac{3w_N\sigma_N}{R_N},
\end{eqnarray}
where the variables with index of $N$ denote the quantities in the nuclear pasta, and $f_3(w)\equiv 1-3w^{1/3}/2+w/2$ \cite{rev1}. In a similar way, with respect to the HQ pasta, one can write down 
\begin{eqnarray}
  \frac{E_{\rm Coul}^Q}{V_Q} &=& \frac{1}{w_Q}\frac{4\pi}{5}\left(Y_e^Q-Y_p^{Q}\right)^2\left(n_B^Q\right)^2e^2R_Q^2f_3(w_Q), 
     \label{eq:a3} \\
  \frac{E_{\rm surf}^Q}{V_Q} &=& \frac{3w_Q\sigma_Q}{R_Q},
\end{eqnarray}
where the variables with index of $Q$ denote the quantities in the HQ pasta \cite{rev1}. We remark that we adopt the total charge neutrality to derive Eqs. (\ref{eq:a1}) and (\ref{eq:a3}). The optimal value of $R$ is determined by the ``virial theorem" in each phase, i.e., $E_{\rm surf}= 2E_{\rm Coul}$. Then, one can obtain
\begin{eqnarray}
   \left(\frac{R_Q}{R_N}\right)^3 &=& \frac{\sigma_Q}{\sigma_N}\left(\frac{n_B^N}{n_B^Q}\right)^2\left(\frac{Y_e^N}{Y_e^Q-Y_p^Q}\right)^2 \left(\frac{w_Q}{w_N}\right)^2\frac{f_3(w_N)}{f_3(w_Q)} \nonumber \\
   &\simeq& \frac{\sigma_Q}{\sigma_N}\left(\frac{n_B^N}{n_B^Q}\right)^2\left(\frac{Y_e^N}{Y_p^Q}\right)^2 \left(\frac{w_Q}{w_N}\right)^2\frac{f_3(w_N)}{f_3(w_Q)},
\end{eqnarray}
where we assume that $Y_e^Q \ll Y_p^Q$ \cite{Maruyama2007,Yasutake2009}. Supposing that the total charge of droplet in the nuclear pasta, $Z_N$, and in the HQ pasta, $Z_Q$, could be expressed as $Z_N\propto 4\pi R_N^3n_p^N/3$ and $Z_Q\propto 4\pi R_Q^3n_q/3$, one can obtain the relationship as
\begin{equation}
   \frac{Z_Q}{Z_N} \sim \frac{n_qR_Q^3}{n_p^NR_N^3}.
\end{equation}
Additionally, $n_q$ and $n_p^N$ can be written as
\begin{eqnarray}
   n_q &\simeq& \frac{w_Q-1}{w_Q}n_p^Q = \frac{w_Q-1}{w_Q}Y_p^Qn_B^Q, \\
   n_p^N &=& Y_p^N n_B^N = \frac{Y_e^N}{w_N}n_B^N.
\end{eqnarray}
Thus, one can get 
\begin{equation}
   \frac{Z_Q}{Z_N} \sim \frac{\sigma_Q n_B^N Y_e^N w_Q (w_Q-1)f_3(w_N)}{\sigma_N n_B^Q Y_p^Q w_N f_3(w_Q)}. \label{eq:ZZZ}
\end{equation}
Since $\sigma_N$ is typically $\sim$1 MeV fm$^{-2}$ and we adopt $\sigma_Q=(10-40)$ MeV fm$^{-2}$, $\sigma_Q/\sigma_N$ becomes $\sim 10-40$. In addition, according to Refs. \cite{Maruyama2007,DH}, we can adopt $Y_e^N=0.161$, and $Y_p^Q=0.250$ at $n_B^N\simeq 0.04$ fm$^{-2}$ and $n_B^Q=0.4$ fm$^{-2}$. For this case, one can get $f_3(w_Q)\sim 0.106$ and $f_3(w_N)\sim 0.471$ with $w_Q\sim 0.375$ and $w_N\sim 0.051$. Then, one can obtain that $|Z_Q/Z_N|\sim 13.3-53.0$, which is in good agreement with the values in Fig. \ref{fig:ZZ}. At last, we should remark that the estimation in this appendix is done only with using the relationship between the Coulomb and the surface energies, i.e., the result is independently of the component in the pasta phase. So, even if the HQ pasta would appear in higher density region with another EOS, one may be able to estimate the charge of droplet and shear modulus via Eq. (\ref{eq:ZZZ}) similarly.







\end{document}